\documentclass[twocolumn,trackchanges]{aastex62}
\usepackage{amsmath}
\hypersetup{
	colorlinks	= true,
	linkcolor	= red,
	urlcolor	= cyan,
	citecolor	= blue
}

\usepackage{amssymb}
\usepackage{soul}
\usepackage{lipsum}
\usepackage{multirow}

\graphicspath{{./images/}}

\newcommand{\exorelr}{\mbox{\textsc{ExoReL$^\Re$}}}

\turnoffediting

\shortauthors{Author1 \& Author2}

\usepackage{fancyhdr}
\begin{document}

 \title{\Large \textbf{The Detectability of CH$_4$/CO$_2$/CO and N$_2$O Biosignatures through Reflection Spectroscopy of Terrestrial Exoplanets}}
	\correspondingauthor{Armen Tokadjian}
	\email{armen.tokadjian@jpl.nasa.gov}
	
	\author[0000-0002-4675-9069]{Armen Tokadjian}
	\affiliation{Jet Propulsion Laboratory, California Institute of Technology, Pasadena, CA 91011, USA}
	
	\author[0000-0003-2215-8485]{Renyu Hu}
	\affiliation{Jet Propulsion Laboratory, California Institute of Technology, Pasadena, CA 91011, USA}
    \affiliation{Division of Geological and Planetary Sciences, California Institute of Technology, Pasadena, CA 91125, USA}
    
    \author[0000-0002-1830-8260]{Mario Damiano}
	\affiliation{Jet Propulsion Laboratory, California Institute of Technology, Pasadena, CA 91011, USA}

	\begin{abstract}
The chemical makeup of Earth's atmosphere during the Archean (4 Ga-2.5~Ga) and Proterozoic eon (2.5 Ga-0.5~Ga) contrast considerably with the present-day: the Archean was rich in carbon dioxide and methane and the Proterozoic had potentially higher amounts of nitrous oxide. CO$_2$ and CH$_4$ in an Archean Earth analog may be a compelling biosignature because their coexistence implies methane replenishment at rates unlikely to be abiotic. However, CH$_4$ can also be produced through geological processes, and setting constraints on volcanic molecules like CO may help address this ambiguity. N$_2$O in a Proterozoic Earth analog may be evidence of life because N$_2$O production on Earth is mostly biological. Motivated by these ideas, we use the code $\exorelr$ to generate forward models and simulate spectral retrievals of an Archean and Proterozoic Earth-like planet to determine the detectability of CH$_4$, CO$_2$, CO, and N$_2$O in their reflected light spectrum for wavelength range 0.25-1.8~$\mu$m. We show that it is challenging to detect CO in an Archean atmosphere for volume mixing ratio (VMR) $\leq$~10\%, but CH$_4$ is readily detectable for both the full wavelength span and truncated ranges cut at 1.7$\mu$m and 1.6$\mu$m, although for the latter two cases the dominant gas of the atmosphere is misidentified. Meanwhile, N$_2$O in a Proterozoic atmosphere is detectable for VMR=$10^{-3}$ and long wavelength cutoff $\geq~1.4\mu$m, but undetectable for VMR~$\leq~10^{-4}$ . The results presented here will be useful for the strategic design of the future Habitable Worlds Observatory and the components needed to potentially distinguish between inhabited and lifeless planets.
	\end{abstract}
	
	
	\section{Introduction} \label{sec:intro}
	
	The Astro2020 Decadal Survey has recommended a large space telescope primarily aimed at characterizing small, potentially habitable exoplanets through direct imaging \citep{national2021pathways}. The Habitable Worlds Observatory (HWO) is a mission concept that was created in response to this recommendation which will aim to detect biosignatures on such planets, a major step forward in our search for extrasolar life. A number of studies have summarized the plausibility and observational implications of various biosignatures \citep{schwieterman2018exoplanet,fujii2018,catling2018} with many focused on the role of oxygen in the form of oxygen gas and ozone \citep{leger1993,desmarais2002,meadows2018}. Given that the current Earth hosts an atmosphere with plenty of O$_2$ and O$_3$ that may exhibit strong spectral features, it is no surprise that these molecules are the main component of biosignature studies and likely a smoking gun in detecting life on another planet. However, since Earth has hosted life for billions of years, its modern state represents only a fraction of its history and its current atmosphere represents only a single case of what we may detect in an inhabited planet.

    The Archean eon (4 Ga - 2.5 Ga) features some of the earliest stages of life on Earth \citep{ohtomo2014}. The atmosphere during this time was quite different than it is today, dominated by nitrogen gas, carbon dioxide, and methane, with little to no oxygen present \citep{catling2020archean}. In Archean Earth, biogenic methanogenesis was the primary producer of CH$_4$, which is why CH$_4$ has been suggested as a biosignature for many decades \citep{sagan1993,thompson2022}. The chemical disequilibrium between CH$_4$ and CO$_2$ strengthens the case for these molecules as a biosignature pair not only because the coexistence of such a redox pair could be evidence of biological metabolism, but also because the short photochemical lifetime of CH$_4$ implies replenishment at rates that are unlikely to be abiotic \citep{krissansen2018dis}. However, both of these molecules can be produced through volcanism, and CH$_4$ is also a product of serpentinization \citep{schindler2000}. These non-biological sources create ambiguity and false positive scenarios for the interpretation of biosignatures. Meanwhile, volcanoes also emit CO, with a CO/CH$_4$ flux ratio typically $>$ 0.1 while methanogenesis does not produce CO \citep{burton2013}. Therefore, constraints on atmospheric CO could be used to break (at least some of) the degeneracy \citep{sholes2019,wogan2020}. In this study, we evaluate the potential to detect CO$_2$, CH$_4$, and CO gases through reflection spectroscopy with the currently outlined capabilities of HWO, to determine the extent in which we can distinguish between a lifeless and potentially inhabited planet with anoxic atmospheres.

    The Proterozoic eon (2.5 Ga - 0.5 Ga) followed the Archean and is marked by increasing levels of oxygen and the Great Oxidation Event \citep{lyons2014}. The presence of oxygen allowed for the growth of organisms crucial to the nitrogen cycle, and its yet limited amount contributed to the incomplete denitrifying step and a buildup of N$_2$O in the atmosphere \citep{chen2015}, a compelling biosignature due to its strong spectral features in the near-IR and limited abiotic sources (e.g., \citealp{rauer2011}). For this reason, and due to the likelihood of higher N$_2$O mixing ratios in the Proterozoic Earth compared to other eons \citep{buick2007}, we simulate the atmosphere of a Proterozoic Earth-like planet with varying amounts of N$_2$O to investigate the detectability of this gas in this scenario, and its potential as an alternative biosignature.

    We use the code \exorelr, an atmospheric model generator for reflection spectra and Bayesian retrieval framework \citep{Damiano2020a}, to perform all atmosphere simulations and retrievals. In \exorelr, a cloud and radiative transfer model determines the wavelength dependence of the flux ratio of light from planet to light from star. Cloud properties such as depth, thickness, and chemical identity are fit to the model and influence the retrieved non-uniform volume mixing ratio (VMR) of the condensable molecular species. We make use of a newer version of the code with specific enhancements highlighted in the text. Here we simulate several Archean Earth-like scenarios with varying amounts of CO and determine the constrained VMR of molecules in the atmosphere. We also simulate two Proterozoic Earth-like scenarios, each with a different input VMR for N$_2$O. In both the Archean and Proterozoic cases, we use a wavelength range 0.25 - 1.8 microns, covering the near-UV, visible, and near-IR with resolutions of 7, 140, and 70, but also investigate alternative wavelength ranges for each case. We also simulate a scenario for an Archean Earth-like planet with a paucity of CO$_2$, to investigate the extent to which the spectral signature of CO$_2$ masks the signal of CO.

    This paper is organized as follows: in Section~\ref{sec:methods}, we outline the relevance of CH$_4$ and N$_2$O biosignatures in an Archean and Proterozoic Earth context and describe our retrieval setup. In Section~\ref{sec:results}, we summarize the results of our forward models and retrievals and determine the potential of detecting certain molecular species. We discuss these results and their implications on future observations in Section~\ref{sec:discussion}, and end with a conclusion in Section~\ref{sec:conclusion}.

	\section{Methods} \label{sec:methods}

        In this section, we first introduce the relevant background and details for the biosignatures of interest in this paper. We then describe the atmospheric modeling and retrieval setup used to study the reflection spectrum and detectability of these molecules in an exo-Earth context.

       \subsection{Biosignatures}\label{subsec:biosignatures}

       As a major atmospheric component, of predominantly biological origin, and with few sources of abiotic production, oxygen is perhaps the most robust biosignature in a modern Earth-like environment \citep{harman2015}. However, recent studies have shed light on numerous pathways of oxygen formation and buildup from geological and photochemical pathways, including ocean vaporization and photolysis of H$_2$O \citep{luger2015} and the destruction of CO$_2$ from UV rays \citep{gao2015}. Thus, due to the possibility of oxygen detection being a false positive in the context of life, it is important to consider alternative biosignatures in addition to O$_2$ and O$_3$. Here we focus on two cases: CH$_4$ in an atmosphere resembling Archean Earth and N$_2$O in a Proterozoic Earthlike environment. Although these molecules are not free from the risk of false positives, incorporating a variety of potential biosignatures in our studies and considering planetary contexts that differ from modern Earth will paint a more complete picture of the potential to detect life on another planet.

	\subsubsection{CH$_4$ in Archean Earth} \label{subsubsec:archean}
	
    The Earth's atmosphere during the Archean eon was very different from today. Marked mainly by the lack of oxygen, the atmospheric makeup constituted mostly of nitrogen, carbon dioxide, and methane. The coexistence of the CO$_2$/CH$_4$ pair which are effectively on opposite ends of carbon's redox spectrum implies a strong disequilibrium in the atmosphere and therefore a replenishing source for the CH$_4$ molecule \citep{thompson2022}. Given its short photochemical lifetime in an oxidizing environment, CH$_4$ must have been produced in the Archean Earth at large fluxes, which make it likely to have a biogenic origin. In fact, the primary source of CH$_4$ during the Archean eon was methanogenesis, a process in which CH$_4$ is emitted as a waste product during microbial respiration \citep{catling2001}. The total surface flux of CH$_4$ into the atmosphere during the Archean may have exceeded 50 Tmol yr$^{-1}$, resulting in a volume mixing ratio (VMR) of 0.01.

    CH$_4$ can be produced abiotically in various ways; the two most prominent are volcanism/outgassing and water-rock reactions, including serpentinization. Thus, there is a risk that the detected CH$_4$ on an exoplanet comes from geothermal systems rather than life, i.e., a false positive scenario. Compared to the total estimated global flux in the Archean, however, the abiotic contribution of CH$_4$ was relatively minor, less than 1 Tmol yr$^{-1}$ \citep{kasting2005}. Indeed, \citealt{krissansen2018dis} argues that it is unlikely for these sources to maintain large enough fluxes to contribute significantly to global CH$_4$ amounts. However, recent studies suggest that high CH$_4$ emission is possible through serpentinization and the subsequent reaction between H$_2$ and residual CO$_2$ on early Earth \citep{miyazaki2022}; and in the event that such large fluxes do exist in an exoplanetary context, other identifiable features may be required to reveal the abiotic origin of the CH$_4$.

    One feature that could help identify volcanic sources of CH$_4$ is the co-presence of CO in the atmosphere, as volcanic outgassing typically produces more CO than CH$_4$  \citep{thompson2022}. In addition, CO is a source of free energy and is consumed by life on Earth \citep{ragsdale2004}. For these reasons, the detection of CO makes CH$_4$ more likely abiotic. In other words, detecting CH$_4$ and CO$_2$ with an absence of CO on an exoplanet could be a strong biosignature, whereas detecting all three molecules together is likely an indication of active volcanism and thus a false positive.

    To quantify the amount of CO relative to CH$_4$ needed for CO to be an ``antibiosignature'' marker, coupled ecological-atmospheric models have been developed that track the CO/CH$_4$ ratio as a function of volcanic H$_2$ flux \citep{kharecha2005,schweiterman2019}. In \citealt{thompson2022}, it is shown that for Archean volcanism, a ratio of 4 - 10 is indicative of an abiotic environment where CH$_4$ is produced volcanically whereas a ratio of $\lesssim 0.1$ signifies an ecosystem that includes methanogens, CO metabolizing organisms, and organic matter fermentation. This means that a planet with atmospheric CO abundances comparable to or larger than CH$_4$ levels is likely to have produced that CH$_4$ abiotically, and a detection of such a case should be interpreted as a potential false positive of the CH$_4$ biosignature. In this study, we take VMR$_{\textrm{CH}_4}=0.01$ which represents an upper limit of CH$_4$ in the Archean atmosphere and consider VMR$_{\textrm{CO}}=0.01$, $0.05$, and $0.1$, corresponding to CO/CH$_4$ ratios of 1, 5, and 10 respectively.

    It has been shown that a CO-dominated atmosphere can be ruled out in an Earth-analog planet for reflected light observations with SNR $\geq$ 20 and wavelength coverage extending beyond 1.6$\mu$m \citep{hall2023}. Although we also study CO detectability, the question addressed here is entirely different: while ruling out CO as the dominant atmospheric species is necessary to exclude certain false positive scenarios for O$_2$ biosignatures \citep{meadows2018exoplanet}, the volcanic CO considered in this work does not accumulate to become the dominant gas. Nevertheless, we also take an SNR of 20 and experiment with the effect of long wavelength cutoff on retrieved molecular abundances.

    Due to many sources of uncertainty, including the geothermal systems on exoplanets and the ratio of CO to CH$_4$ produced abiotically in such worlds, we briefly describe alternative methods of distinguishing between methane on an inhabited versus lifeless planet. One method is to identify the isotope ratio of carbon: life on Earth preferentially prefers $^{12}$C over other isotopes, such that biologically produced methane will predominantly bear the $^{12}$C isotope, whereas volcanically sourced methane will have comparatively smaller ratios of $^{12}$C/$^{13}$C \citep{meister2019}. However, there is no guarantee that life on other planets will be similar to Earth; the enzymes and metabolisms that evolved on exoplanet surfaces may be quite different. In addition, it is challenging to detect such isotopic differences in habitable exoplanets with current instrumentation. Thus, CO remains the most reasonable candidate for unraveling the origin of CH$_4$ and identifying false positives.
    
    \cite{damiano2022} have already studied the performance of \exorelr on retrieving molecular abundances in an Archean Earth-like atmosphere. It was found that CO$_2$ and CH$_4$ mixing ratios were correctly constrained with $>2\sigma$ confidence. In this paper, we expand upon these results by including CO in addition to all of the molecules and cloud properties presented there and incorporate additional updates such as cloud fraction, updated Rayleigh scattering, and noise realization. In this way, we investigate the detectability of the 1.55$\mu$m CO feature in an Archean atmosphere. \cite{damiano2022} also test the effect of cutting the long wavelength range at 1$\mu$m and found that the near-IR is crucial to constrain CH$_4$. As discussed above, we expand upon this experiment by running retrievals of the Archean scenario for wavelength ranges up to 1.6 and 1.7 micron.
	
	\subsubsection{N$_2$O in the Proterozoic Earth} \label{subsubsec:proterozoic}

    As the Archean eon transitioned into the Proterozoic, atmospheric oxygen levels began to rise, continents began to take shape, and the first eukaryotic organisms evolved \citep{fakhraee2023}. Among other natural phenomena, the nitrogen cycle is sensitive to the availability of oxygen, such that the Proterozoic atmosphere experienced more complex changes than simply the rise of oxygen. For instance, environmental conditions can alter the amount of emitted N$_2$O, a molecule of interest due to its potential as a strong biosignature. In fact, microbes participating in the nitrogen cycle were the predominant producers of N$_2$O \citep{schwieterman2018exoplanet}. This production of N$_2$O can follow many pathways. The primary source is the denitrification step: as nitrates are converted to molecular nitrogen, an incomplete reaction can lead to buildup of the intermediate N$_2$O molecule \citep{quick2019}. Other pathways include hydroxylamine oxidation by bacteria and dissimilatory nitrate reduction to ammonium (DNRA) by bacteria/fungi \citep{pinto2021}.
 
     N$_2$O is an attractive biosignature because there are few abiotic sources. On early Earth, perhaps the most significant contributor was chemodenitrification in hypersaline lakes, a process in which nitrates are abiotically reduced to N$_2$O by ferrous ions \citep{samarkin2010}. However, it is unlikely for this effect to produce N$_2$O at rates significant enough to cause a false positive because of its reliance on disequilibrium between a nitrate rich and reducing ocean \citep{schwieterman2022}. Also, \citealt{Hu2019} showed that dehydrative dimerization of HNO would produce less N$_2$O than Earth's biosphere. Other abiotic sources include lightning which contributes only 0.002\% of atmospheric N$_2$O \citep{schumann2007} and extreme ultraviolet radiation which will produce far more detectable nitrogen oxides than N$_2$O \citep{schwieterman2018exoplanet}.

    We choose the Proterozoic Earth context because of the advantages it provides over other historical eons in terms of producing and maintaining an N$_2$O-detectable atmosphere. In contrast to modern Earth, the Proterozoic atmosphere contained little oxygen, and an anoxic and sulfur-rich ocean limits the amount of copper catalysts needed for the reduction of nitrates to nitrogen gas \citep{knowles1982}. This effectively hinders the denitrification step in the nitrogen cycle and results in a greater abundance of the intermediate product, N$_2$O. The Archean Earth contained little to no oxygen, which may have had a similar effect on the nitrogen cycle. However, the most significant sink of N$_2$O is photolysis by UV photons, where the presence of oxygen and notably ozone can mitigate this effect \citep{airapetian2016}. In fact, photochemical modeling has shown that without a protective ozone layer, marine abiotic N$_2$O emissions would sustain less than 10$^{-8}$ atmospheric N$_2$O VMR. \citep{buessecker2022}. Thus, it is this intermediate amount of oxygen that enables an increase in the production and longevity of the N$_2$O molecule.

    For these reasons, Proterozoic Earth may have contained N$_2$O in abundances orders of magnitude larger than in Archean or Modern Earth, the latter of which has a mixing ratio $\leq$ 10$^{-6}$ \citep{lemke2007}. Geochemical modeling has shown that N$_2$O fluxes just one or two orders of magnitude larger than today's rate of 0.4 Tmol yr$^{-1}$ could lead to atmospheric mixing ratios of more than 10$^{-5}$ and even up to 10$^{-4}$ \citep{schwieterman2022}. Since UV photons play the role of a major N$_2$O sink, these abundances can be further enhanced if we consider K and M stars whose lower UV fluxes can help preserve atmospheric N$_2$O \citep{segura2005}. In fact, a Modern Earthlike planet around a K6 type star can accumulate a mixing ratio of up to 10$^{-5}$, and a Proterozoic Earthlike planet with greater N$_2$O flux can host an upper limit of 10$^{-3}$. Thus, in this work we consider the two cases of N$_2$O mixing ratios, one corresponding to an upper limit for an Earthlike planet around a G-type star, VMR$_{N_2O}=10^{-4}$ and one for the same kind of planet around a K-type star, VMR$_{N_2O}=10^{-3}$ and determine its detectability in a Proterozoic context. It is reasonable to consider exo-Earths around K-type stars because they make up nearly 25\% of the current HWO target star list \citep{mamajek2024}.

    \subsection{Retrieval Setup}
        \label{subsec:setup}
    We use the Bayesian retrieval code \exorelr as described in \citealt{Damiano2020a}: planet reflection spectra are simulated using a radiative transfer model, which is fed into a nested sampling retrieval framework to fit the parameters of choice. We invoke version 2.3.5, with updates to molecular and Rayleigh contributions, cloud fraction possibility, noise realization, and fitting parameterization. This includes recent updates to the N$_2$O opacity line list which incorporates crucial features for $\lambda < 1.2\mu$m (HITEMP2019; \citealp{rothman2010,hargreaves2019}), partial cloud coverage, and an adaptive grid with 100 layers (Damiano et al., in prep). We choose to fit for partial pressures, and the free parameters and their prior ranges are outlined in Table~\ref{free_param_table}.  

       \begin{deluxetable}{cclc}
		\tablecaption{Free Parameters and Prior Ranges}
		\tablehead{
			\colhead{Free Parameter} & \colhead{Symbol} & \colhead{Range} & \colhead{Type}}
		\startdata
		\underline{Clouds} & & & \\
         Cloud Top& $P_{\mathrm{top},H_2O}$ & [2 , 7] Pa & Log-uniform\\
         Cloud Thickness& $D_{H_2O}$ & [2 , 7] Pa & Log-uniform\\
         Condensation Ratio& $CR_{H_2O}$ & [-7 , 0] & Log-uniform\\
         \underline{Molecules} & & & \\
         H$_2$O Partial Pressure & H$_2$O & [-2 , 7] Pa & Log-uniform\\
         CH$_4$ Partial Pressure & CH$_4$ & [-2 , 7] Pa & Log-uniform\\
         CO$_2$ Partial Pressure & CO$_2$ & [-2 , 7] Pa & Log-uniform\\
         CO Partial Pressure & CO & [-2 , 7] Pa & Log-uniform\\
         N$_2$O Partial Pressure & N$_2$O & [-2 , 7] Pa & Log-uniform\\
         O$_2$ Partial Pressure & O$_2$ & [-2 , 7] Pa & Log-uniform\\
         \underline{Misc.} & & & \\
         Surface Albedo$^a$ & $A_g$ & [-2 , 0] & Log-uniform\\
         Planet Radius & $R_p$ & [0.5 , 10] $R_{\oplus}$ & Linear-uniform
		\enddata
       \label{free_param_table}
       \tablecomments{ $^a$This applies to retrievals on Proterozoic Earth. For Archean Earth, we set $A_g=0.2$ }
      \end{deluxetable}
    
    Spectra are generated for the Archean and Proterozoic case using parameters from \citealt{krissansen2018dis}, \citealt{damiano2022}, and \citealt{damiano2023}, with the additions of CO and N$_2$O as appropriate, summarized together with the retrieval results in Table~\ref{archean_table} and Tables \ref{proterozoic_table_-3} and \ref{proterozoic_table_-4}. Included in the spectra is Gaussian noise applied to a photon noise model scaled to an SNR of 20 at 0.75$\mu$m. The wavelength range is 0.25$\mu$m-1.8$\mu$m, covering the near UV, optical, and near IR, and we use resolutions of $R=7$, $R=140$, and $R=70$, respectively. These are values considered for the HabEx and LUVOIR concept, precursors of the Habitable Worlds Observatory \citep{luvoir2019,Gaudi2020}. In the following section, we describe the results of our spectral retrievals.

	\section{Results} \label{sec:results}

	\subsection{CH$_4$/CO$_2$ in Archean Earth Context} 
	\label{subsec:archeanr}
\begin{figure*}
   \epsscale{1.2}
   \plotone{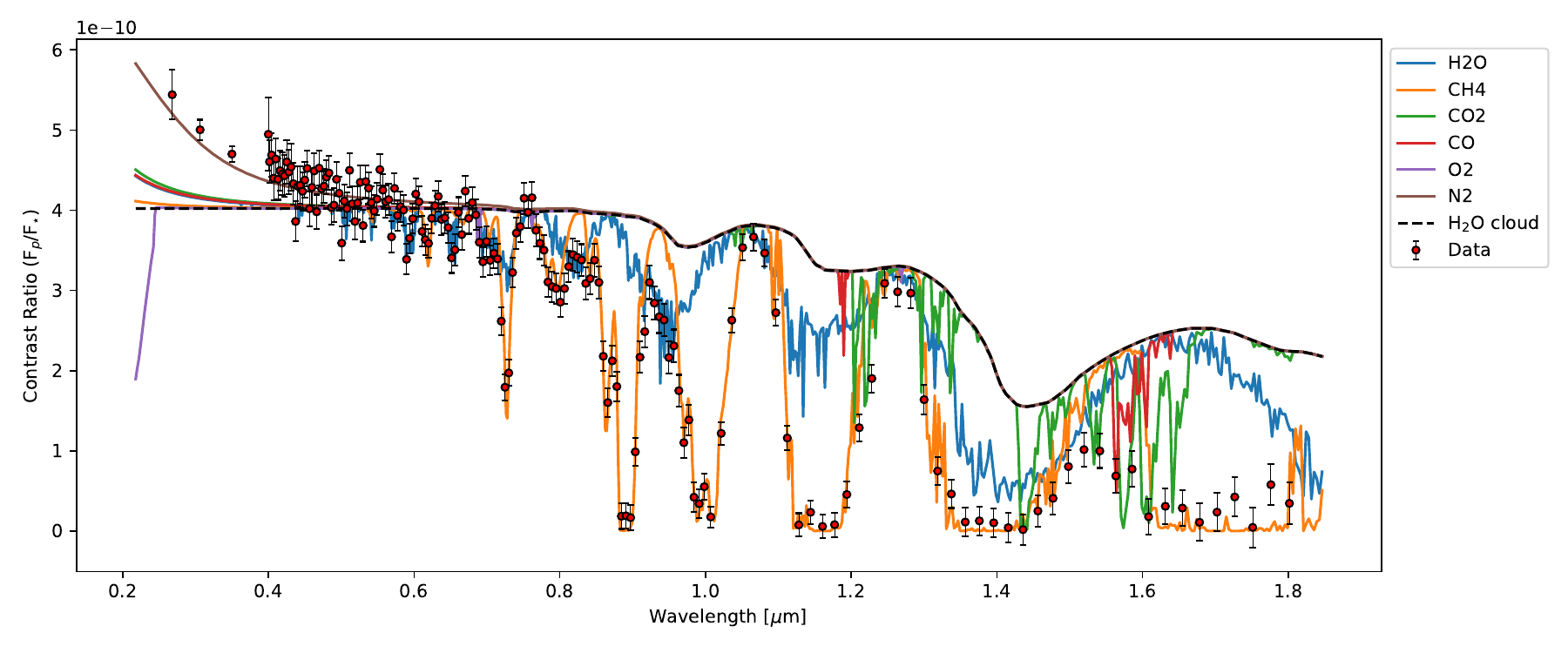}
    \caption{Flux contrast ratio of an Archean Earthlike planet with CO/CH$_4$ = 10 as a function of wavelength. The input forward model is shown by the red circles, and the retrieval result is overlain and separated by molecule to show the individual molecular contributions. }
    \label{Archean_contr_10x}
    \epsscale{1.0}
    \end{figure*}

\begin{figure}
   \epsscale{1.2}
   \plotone{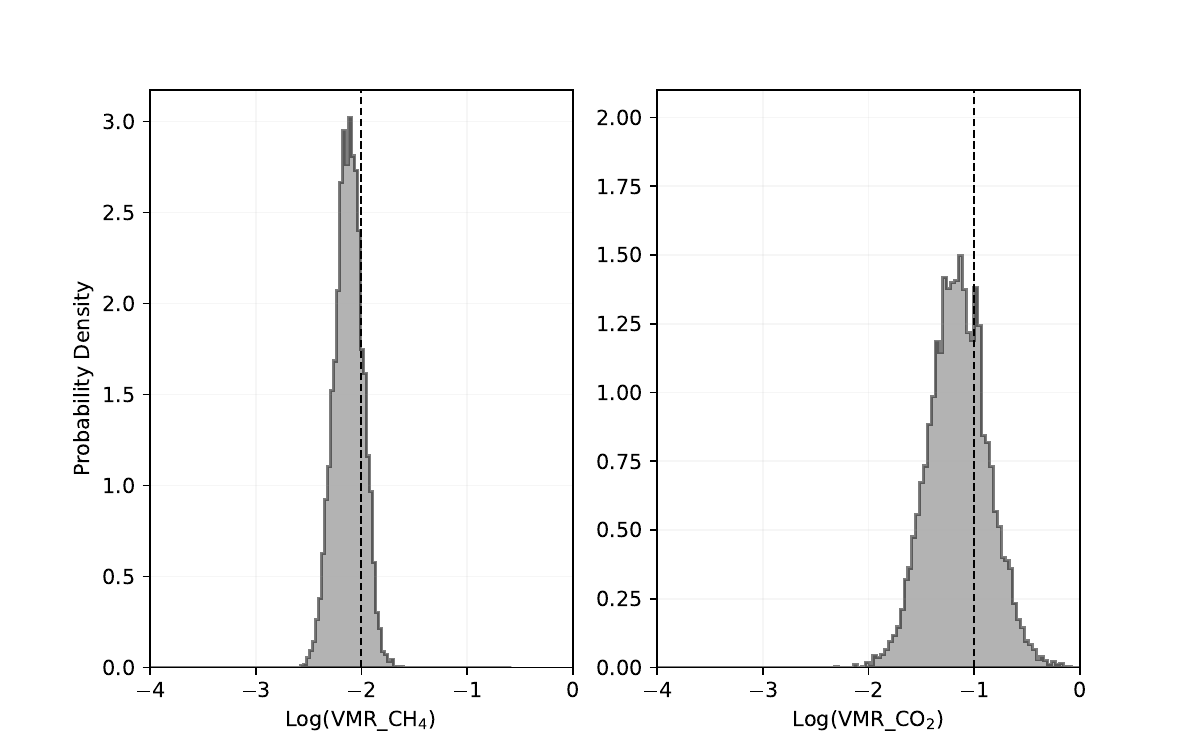}
    \caption{Posterior solution curves for the mixing ratios of  CH$_4$ (left) and CO$_2$ (right) for the same case shown in Figure~\ref{Archean_contr_10x}. The dashed vertical lines represent the input value to the forward model, which agree well with the posterior results.}
    \label{CO2_CH4_post_10x}
    \epsscale{1.0}
    \end{figure}

	To model the Archean Earth atmosphere, we use a 50-50 cloud coverage scenario and set the surface albedo to a constant 0.2, consistent with previous works (e.g., \citealt{damiano2023}). Because CO features in the wavelength range presented here are few and weak, making these assumptions will potentially improve our ability to retrieve the correct CO abundance. In this way, our results for the Archean Earth represent a somewhat optimistic outlook on CH$_4$/CO$_2$/CO retrieval. Figure~\ref{Archean_contr_10x} shows the planet-to-star flux ratio as a function of wavelength for an Archean scenario with CO/CH$_4$ = 10 (red circles) along with the retrieved molecular and cloud contribution (colored curves). The strongest features are due to CH$_4$ (orange), followed by CO$_2$ (green) and H$_2$O (blue). Due to these features, it is unsurprising that we obtain well constrained posterior solutions for the retrieval on CH$_4$ and CO$_2$ abundances (Figure~\ref{CO2_CH4_post_10x}).

\begin{figure}
   \epsscale{1.35}
   \plotone{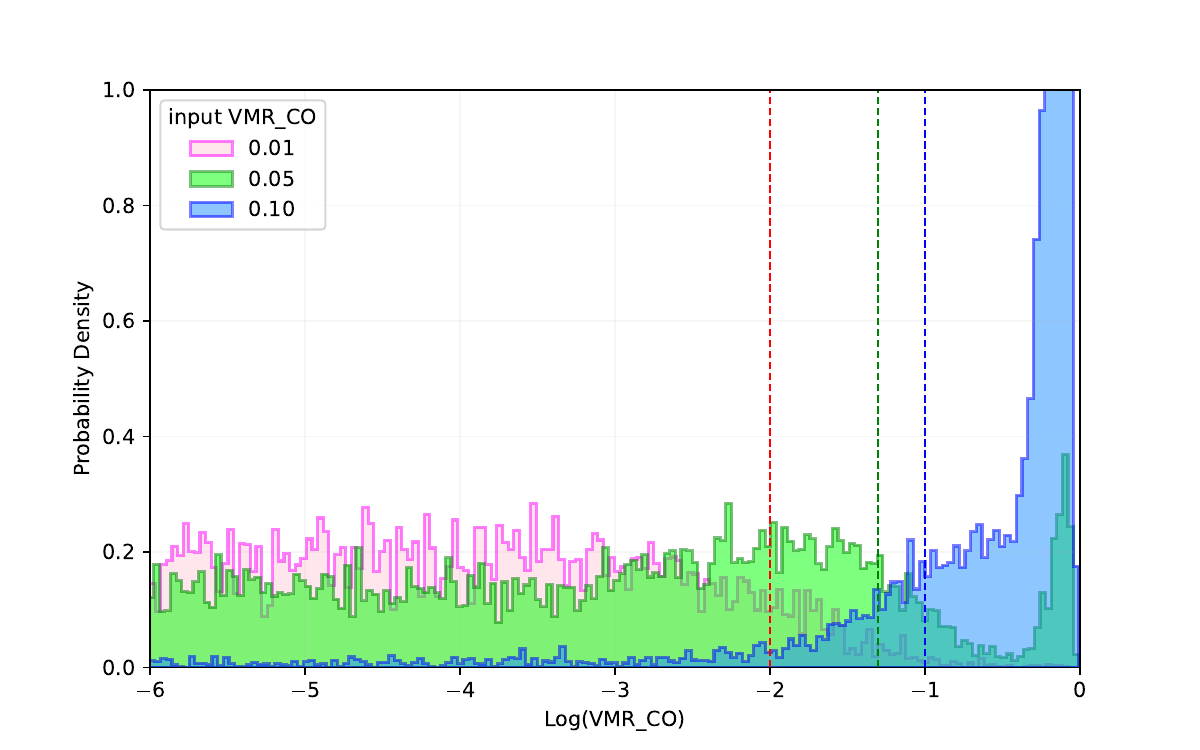}
    \caption{Posterior solution curves for the abundance of CO for CO/CH$_4$ ratios of 1, 5, and 10, corresponding to CO VMR's of 0.01 (pink), 0.05 (green), and 0.1 (blue) respectively. The dashed vertical lines represent the input values of CO, color coded to the corresponding histogram. For small ratios, the CO distribution is wide, signifying a non-detection. For moderate and high ratios, a peak appears for large CO abundance, which is degenerate with N$_2$. Thus, with the current retrieval setup, it will be challenging to constrain CO abundance in an Archean Earth scenario.}
    \label{1x5x10xCOpost}
    \epsscale{1.35}
    \end{figure}

    Even with the maximum amount of CO we consider in this study (VMR$_\mathrm{CO}$ = 0.1), the CO features shown in red in Figure~\ref{Archean_contr_10x} are relatively weak. The main absorption is at 1.55$\mu$m, which is located within a stronger CO$_2$ feature. Thus, we do not attain good constraints on CO abundance for any of the three cases we consider (Figure~\ref{1x5x10xCOpost}). For CO/CH$_4$ = 1, we obtain a wide posterior, signifying a general non-detection of CO. In the CO/CH$_4$ = 5 case, the posterior is again wide, but has a peak at large VMR, representing a CO dominated atmosphere. However, the Bayesian evidence for such an atmosphere is small, and again the retrieval prefers a low amount of CO and a N$_2$ dominated case (See Table~\ref{archean_table}). The CO/CH$_4$ = 10 case no longer leads to a wide posterior but retains the peak at high values, and includes a tail toward low CO abundance, where the 1$\sigma$ and 3$\sigma$ lower limits are -0.88 and -6.74, respectively, for Log(VMR$_{CO}$). These limits are consistent with a non-detection of CO. It is also degenerate with N$_2$: a CO dominated and N$_2$ dominated atmosphere are both valid solutions. Thus, using our current setup, an Archean scenario with a moderate to high amount of CO that corresponds to the abiotic case (CO/CH$_4$ $>$ 3) will not display the necessary features to constrain CO abundances and to rule out potential false positives. Additional studies on the SNR, wavelength range, resolution, and atmospheric scenario needed to constrain CO in a CH$_4$/CO$_2$ environment will be important for identifying false positives.

    The results for the Archean case with variable CO abundance are summarized in Table~\ref{archean_table}, for each of the CO mixing ratios we considered. We report the median value of the posterior distribution and 1$\sigma$ error. In each of the scenarios, there are good fits on the VMR of H$_2$O, CH$_4$, and CO$_2$ and cloud parameters (all within or close to 1$\sigma$). As summarized above, CO has the poorest retrieval results; for Log(VMR$_\mathrm{CO}$) of -2 and -1.3, the retrieval underestimates the amount of CO with large error bars and for Log(VMR$_\mathrm{CO}$) = -1, the retrieval overestimates the CO abundance. For this specific case, the VMR of N$_2$ also has a very wide distribution, signifying the degeneracy between CO and N$_2$. Indeed, even though this case includes a relatively high amount of CO, the CO$_2$ feature at $1.55~\mu$m is much stronger and the retrieval cannot distinguish between N$_2$ and CO Rayleigh scattering. This is consistent with results from \citealt{hall2023}. Note that the constraints on N$_2$ are much more precise and accurate for the other two cases where this degeneracy is not encountered.

\begin{figure}
\epsscale{1.35}
\plotone{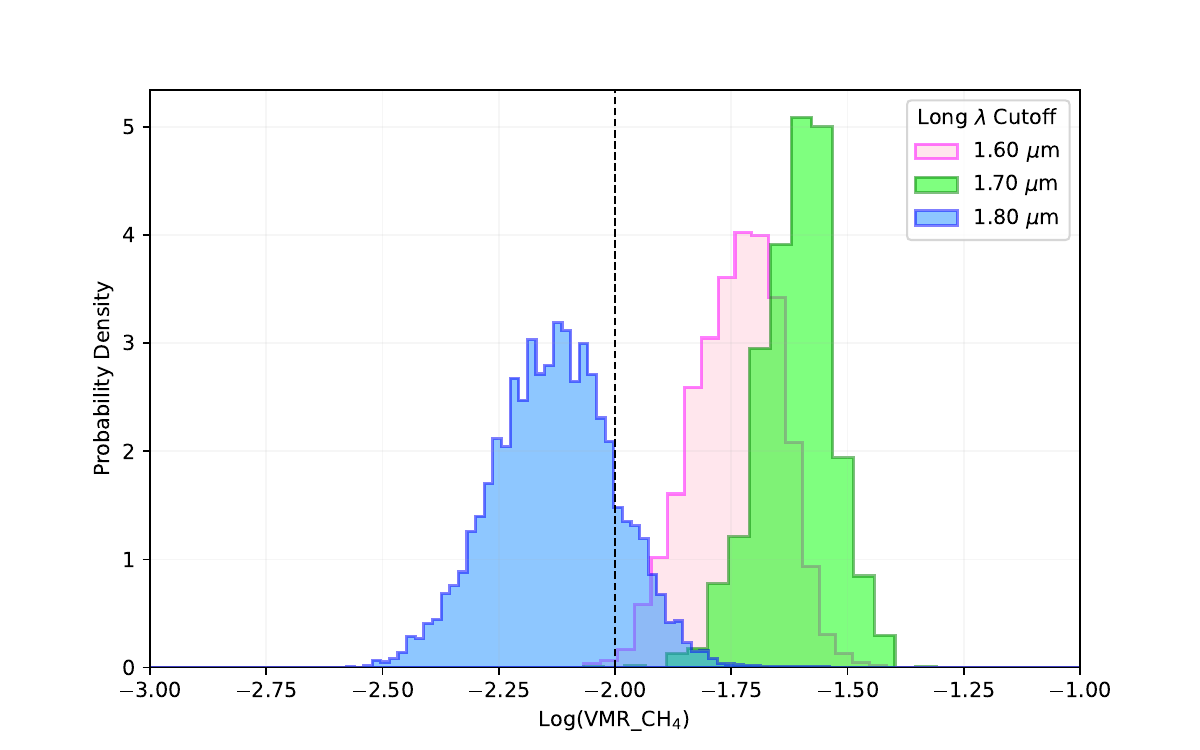}
    \caption{Posterior solution curves for the abundance of CH$_4$ for long wavelength cutoff of 1.60$\mu$m (pink), 1.70$\mu$m (green), and 1.80$\mu$m (blue). The dashed vertical line is the input value of CH$_4$. The full wavelength range results in the best constraint on CH$_4$. The middle and short wavelength cutoff results show a bias towards higher amounts of CH$_4$ due to the different types of atmospheres retrieved in these cases (see text). Nevertheless, all three cases show a good constraint on CH$_4$. }
    \label{archean_wl_post}
    \epsscale{1.35}
    \end{figure}

    For our long wavelength cutoff experiment for the Archean case, we ran retrievals identical to the Log(VMR$_{CO}$)=-2 case above but with maximum wavelengths of 1.7$\mu$m and 1.6$\mu$m to assess the importance of the cutoff on the resulting molecular abundances and cloud parameters. The posterior solutions for CH$_4$ mixing ratio for varying wavelength range are shown for each cutoff in Figure~\ref{archean_wl_post}. The blue solution is for the full wavelength range and is identical to the left plot in Figure~\ref{CO2_CH4_post_10x}: an excellent result whose 1$\sigma$ range covers the exact input value. The green and pink posteriors represent the 1.7$\mu$m and 1.6$\mu$m cutoff, respectively. While still strongly constrained, these are biased towards higher amounts of CH$_4$. The reason for this can be attributed to the type of atmosphere we retrieve in each case. As we cut the long wavelength to 1.7$\mu$m, we lose some of the CH$_4$ anchor and H$_2$O absorption features between 1.7-1.8$\mu$m. This can be seen by the degrading constraint on H$_2$O and condensation ratio in the full posterior results (Table~\ref{archean_table_wl}). Without the strong CH$_4$ features and moderate H$_2$O features in this range, the model allows for a much greater amount of CO$_2$, leading to much less N$_2$ and thus a bias towards higher amounts of H$_2$O and CH$_4$. For a further cut down to 1.6$\mu$m, some CO$_2$ features are lost and instead we obtain a CO dominated atmosphere. This is a similar result as the Log(VMR$_{CO}$)=-1 case where CO and N$_2$ are indistinguishable, leading to CO being the dominant species and a poor constraint on N$_2$. Obtaining three different dominant molecules for the three cutoff cases demonstrates both the importance of selecting a proper wavelength range and the limitations of retrieval results. Despite this fact, we have shown that the posterior CH$_4$ mixing ratio still does not change appreciably when cutting down to smaller long wavelength cutoff, reinforcing the case for CH$_4$ as a strong biosignature candidate.

    \begin{deluxetable}{crrrr}
		\tablecaption{Retrieval Results for Archean Earth Scenario for Varying CO Input Abundance}
		\tablehead{
               & &  \multicolumn{3}{c}{---------Log(VMR$_{CO}$)---------} \\
			\colhead{Parameter} & \colhead{Input} & \colhead{-2} & \colhead{-1.3} & \colhead{-1}
   }
		\startdata
		\underline{Clouds (Log)} & & & \\
         $P_{\mathrm{top},H_2O}$ (Pa) & 4.5  & $4.64^{+0.15}_{-0.15}$ & $4.71^{+0.19}_{-0.20}$ & $4.60^{+0.17}_{-0.16}$\\
         $D_{H_2O}$ (Pa) & 4.0 & $3.74^{+0.23}_{-0.27}$ & $4.00^{+0.20}_{-0.25}$ & $4.01^{+0.25}_{-0.39}$\\
         $CR_{H_2O}$ & $-4.0$ & $-5.05^{+1.24}_{-1.32}$ & $-5.25^{+1.20}_{-1.14}$ & $-4.32^{+1.10}_{-1.77}$\\
         \underline{Mol. (Log)} & & & &\\
         VMR$_{H_2O}$ & $-1.0$ & $-1.01^{+0.17}_{-0.18}$ & $-1.02^{+0.22}_{-0.21}$ & $-0.91^{+0.19}_{-0.20}$\\
         VMR$_{CH_4}$ & $-2.0$ & $-2.13^{+0.13}_{-0.13}$ & $-2.19^{+0.17}_{-0.16}$ & $-2.12^{+0.13}_{-0.14}$\\
         VMR$_{CO_2}$ & $-1.0$ & $-1.42^{+0.26}_{-0.26}$ & $-1.46^{+0.33}_{-0.31}$ & $-1.15^{+0.28}_{-0.27}$\\
         VMR$_{CO}$ & * & $-4.44^{+1.78}_{-1.76}$ & $-3.33^{+1.83}_{-2.49}$ & $-0.16^{+0.07}_{-0.72}$\\
         VMR$_{O_2}$ & $-7.0$ & $-5.13^{+1.40}_{-1.30}$& $-4.32^{+1.56}_{-1.84}$ & $-3.56^{+1.06}_{-2.23}$\\
         VMR$_{N_2}$ & $-0.1$ & $-0.07^{+0.02}_{-0.04}$& $-0.07^{+0.03}_{-0.07}$ & $-2.27^{+2.55}_{-2.83}$\\
         \underline{Misc.} & & & \\
         $R_p$ ($R_{\oplus}$) & 1.0 & $1.00^{+0.01}_{-0.01}$ & $1.00^{+0.00}_{-0.00}$ & $1.00^{+0.00}_{-0.00}$
		\enddata
      \label{archean_table}
       \tablecomments{ * The input VMR for CO is given by the columns. Input values are from \citealt{krissansen2018dis} and \citealt{damiano2022}. }
      \end{deluxetable}      

        \begin{deluxetable}{crrrr}
		\tablecaption{Retrieval Results for Archean Earth Scenario for Varying Wavelength Range}
		\tablehead{
               & & \multicolumn{3}{c}{Long Wavelength Cutoff} \\
			\colhead{Parameter} & \colhead{Input} & \colhead{1.8$\mu$m} & \colhead{1.7$\mu$m} & \colhead{1.6$\mu$m}
   }
		\startdata
		\underline{Clouds (Log)} & & & \\
         $P_{\mathrm{top},H_2O}$ (Pa) & 4.5  & $4.64^{+0.15}_{-0.15}$ & $4.14^{+0.06}_{-0.08}$ & $4.24^{+0.11}_{-0.10}$\\
         $D_{H_2O}$ (Pa) & 4.0 & $3.74^{+0.23}_{-0.27}$ & $3.43^{+0.29}_{-0.41}$ & $3.18^{+0.21}_{-0.17}$\\
         $CR_{H_2O}$ & $-4.0$ & $-5.05^{+1.24}_{-1.32}$ & $-3.81^{+0.99}_{-1.91}$ & $-3.13^{+0.32}_{-0.48}$\\
         \underline{Mol. (Log)} & & & &\\
         VMR$_{H_2O}$ & $-1.0$ & $-1.01^{+0.17}_{-0.18}$ & $-0.52^{+0.15}_{-0.16}$ & $-0.67^{+0.16}_{-0.17}$\\
         VMR$_{CH_4}$ & $-2.0$ & $-2.13^{+0.13}_{-0.13}$ & $-1.60^{+0.06}_{-0.08}$ & $-1.73^{+0.09}_{-0.10}$\\
         VMR$_{CO_2}$ & $-1.0$ & $-1.42^{+0.26}_{-0.26}$ & $-0.20^{+0.07}_{-0.10}$ & $-0.68^{+0.18}_{-0.23}$\\
         VMR$_{CO}$ & $-2.0$ & $-4.44^{+1.78}_{-1.76}$ & $-5.91^{+3.58}_{-4.08}$ & $-0.26^{+0.10}_{-0.16}$\\
         VMR$_{O_2}$ & $-7.0$ & $-5.13^{+1.40}_{-1.30}$& $-7.22^{+2.90}_{-3.04}$ & $-9.69^{+0.97}_{-0.72}$\\  
         VMR$_{N_2}$ & $-0.108$ & $-0.07^{+0.02}_{-0.04}$& $-5.24^{+4.01}_{-4.08}$ & $-8.07^{+3.18}_{-2.13}$\\
         \underline{Misc.} & & & \\
         $R_p (R_{\oplus})$ & $1.0$ & $1.00^{+0.01}_{-0.01}$ & $1.01^{+0.01}_{-0.01}$ & $1.01^{+0.01}_{-0.01}$
		\enddata
      \label{archean_table_wl}
       \tablecomments{Input values from \citealt{krissansen2018dis} and \citealt{damiano2023}. }
      \end{deluxetable} 
    
    
    \subsection{N$_2$O in Proterozoic Earth Context} \label{subsec:proterozoicr}
	
\begin{figure*}
   \epsscale{1.2}
   \plotone{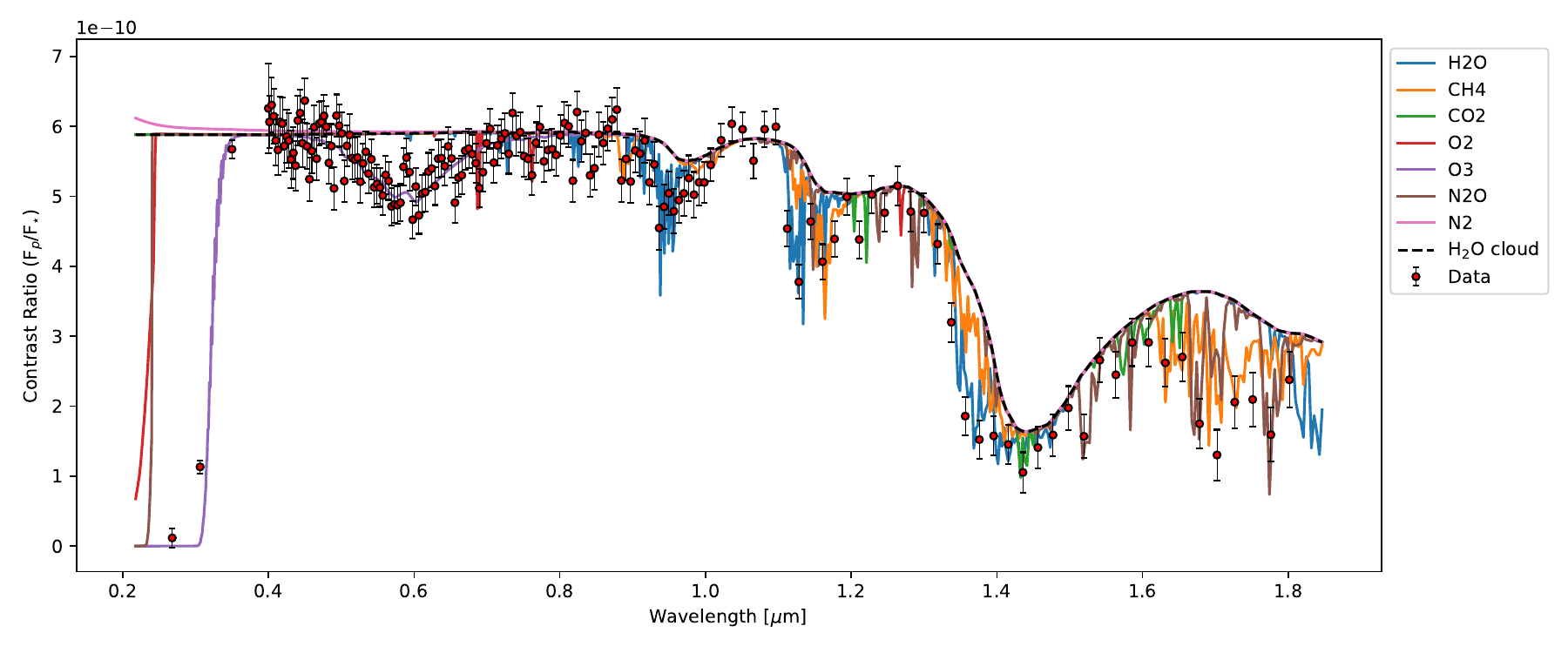}
    \caption{Same as Figure~\ref{Archean_contr_10x}, but for a Proterozoic Earthlike planet with VMR$_{N_2O} = 10^{-3}$. The spectrum does not have any dominating features, but the absorption of numerous molecules are evident. N$_2$O features (brown) can be found at numerous wavelengths past 1$\mu$m, particularly at 1.113, 1.283, 1.383, 1.516, 1.585, 1.671, and 1.777 $\mu$m.}
    \label{Proterozoic_contr}
    \epsscale{1.0}
    \end{figure*}

	In the Proterozoic Earth case, we use a 100\% cloud coverage atmosphere and take surface albedo as a free parameter. This is in contrast to the Archean case and represents a more challenging scenario for molecular abundance detection. Finding good constraints in this setup would reinforce the case for N$_2$O as a strong biosignature. Flux ratio as a function of wavelength for the input simulated data is shown along with molecular contribution retrieval result in Figure~\ref{Proterozoic_contr}. These results make use of the updated opacity tables and cross section data for N$_2$O (HITEMP2019). 

    As outlined above, we simulate Proterozoic Earth-like atmospheres for two N$_2$O mixing ratios, $10^{-3}$ for an Earth-twin orbiting a K-type star and $10^{-4}$ for a G-type host star. We also investigate the effect of long wavelength cutoff on N$_2$O detectability by running three retrievals for each mixing ratio; the wavelength range we employ starts at 0.25$\mu$m and ends at either 1.8$\mu$m, 1.4$\mu$m, or 1.2$\mu$m, motivated by the location of N$_2$O features that are spread throughout the 1.2-1.8$\mu$m range.

\begin{figure}
   \epsscale{1.35}
   \plotone{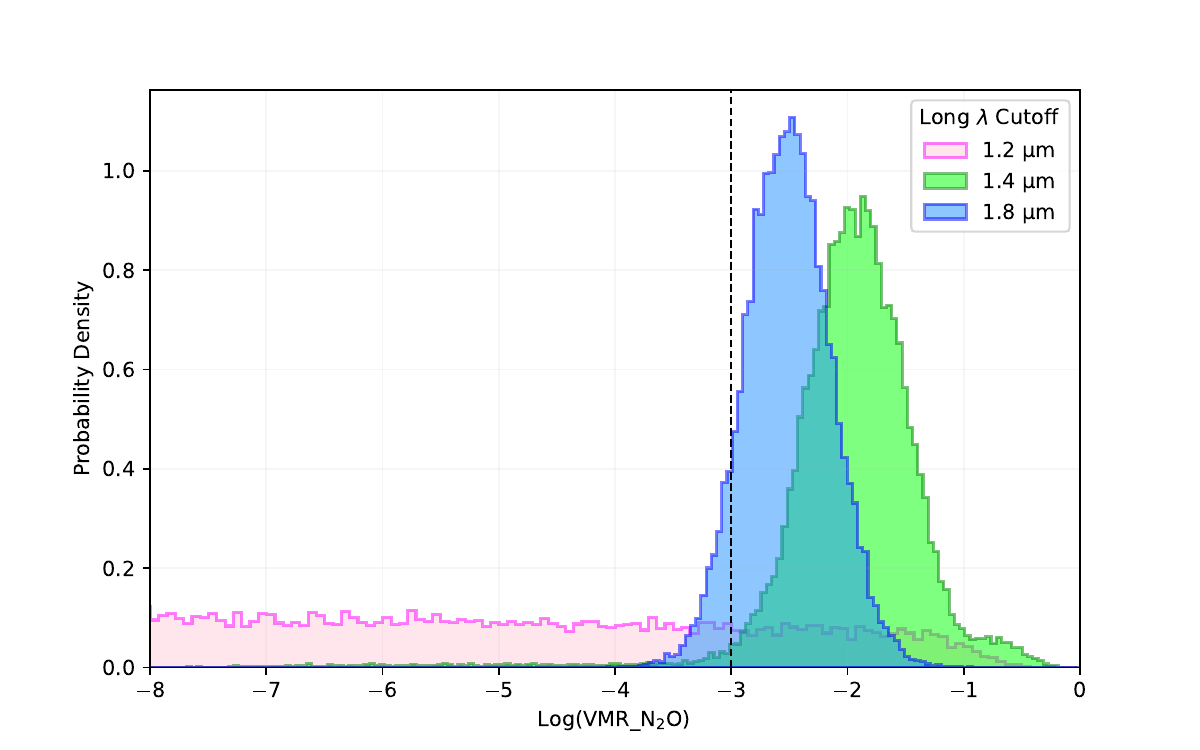}
    \caption{Posterior solution curves for the mixing ratios of  N$_2$O for long wavelength cutoffs at 1.8$\mu$m (blue), 1.4$\mu$m (green), and 1.2$\mu$m (pink). The dashed vertical line is the input value where Log(VMR$_{N_2O})=-3$. A long wavelength cutoff of 1.8$\mu$m and 1.4$\mu$m results in constrained and fairly accurate posteriors, whereas a 1.2$\mu$m cutoff leads to a non-detection of N$_2$O.}
    \label{N2O_post_-3}
    \epsscale{1.35}
    \end{figure}

    The posterior distribution result for the VMR$_{N_2O} = 10^{-3}$ case is shown in Figure~\ref{N2O_post_-3}. A long wavelength cutoff of 1.8$\mu$m (blue) results in a nicely constrained posterior solution for the mixing ratio of N$_2$O. Cutting the wavelength at 1.4$\mu$m (green) preserves the precision of the posterior but with a slightly less accurate result: the probability density peaks at $\sim$ -2 vs -$2.5$ for the previous case. The driving simulated data point here is at 1.383$\mu$m: although the feature is not very deep,  N$_2$O is the strongest absorber at this wavelength for the molecules we fit. Note that this data point is within error of the continuum. This is reflected in the posterior distribution where the green curve tails towards very low amounts of N$_2$O. Finally, we obtain a wide posterior result for a cutoff at 1.2$\mu$m (pink), signifying a non-detection of N$_2$O. This means that for Log(VMR$_{N_2O})=-3$, the wavelength range must extend to $\sim$1.4$\mu$m to reliably detect N$_2$O in a Proterozoic context. The posterior results for this scenario are summarized in Table~\ref{proterozoic_table_-3}. Along with N$_2$O, the posterior results for all other free molecules worsens as the long wavelength cutoff is shrunk. The cloud parameters are also best retrieved for the full wavelength range, but the posteriors distributions do not widen appreciably as we cut to lower wavelengths.

\begin{figure}
   \epsscale{1.35}
   \plotone{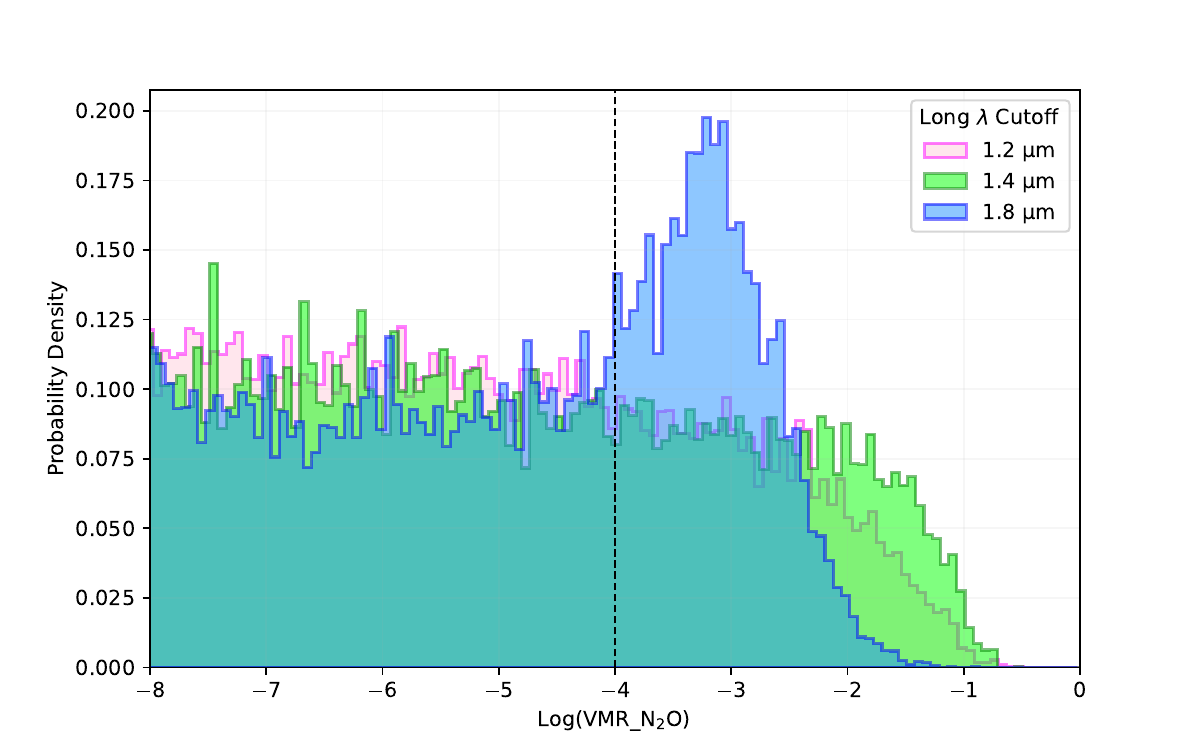}
    \caption{Same as Figure~\ref{N2O_post_-3}, but for an input Log(VMR$_{N_2O})=-4$. For each of the long wavelength cutoffs considered here, we obtain a wide posterior distribution for mixing ratio of N$_2$O, but the case of 1.8$\mu$m cutoff results in a slight peak that is biased towards higher amounts of N$_2$O.}
    \label{N2O_post_-4}
    \epsscale{1.35}
    \end{figure}

    Figure~\ref{N2O_post_-4} shows the posterior distribution results for N$_2$O for the case of VMR$_{N_2O} = 10^{-4}$. Unlike the previous case, we do not obtain well-constrained posterior solutions for any of the three wavelength cutoffs. Both the 1.4$\mu$m and 1.2$\mu$m cases result in peakless and wide posteriors that are consistent with no detection of N$_2$O. The 1.8$\mu$m case has a slight peak at -3, an overestimate of the input value, but is still generally wide and does not well constrain the mixing ratio of N$_2$O. We conclude that VMR$_{N_2O} > 10^{-4}$ is needed to be detected in a Proterozoic Earthlike context, and thus K-type stars would be the best option to search for such planets. Table~\ref{proterozoic_table_-4} summarizes the results for all free parameters: most follow a similar trend as the case in Table~\ref{proterozoic_table_-3}. We note that constraining a N$_2$ dominated atmosphere in this Proterozoic setup depends on the inclusion of features past 1.4$\mu$m due to degeneracy with CO$_2$. In fact, we obtain CO$_2$ as the dominant gas for a cutoff at 1.4$\mu$m and both a CO$_2$ dominated and N$_2$ dominated atmosphere are admissible solutions for a cutoff at 1.2$\mu$m. The surface albedo is also not well constrained for any wavelength cutoff even though the cloud parameters have generally good constraints.

    \begin{deluxetable}{crrrr}
		\tablecaption{Retrieval Results for Proterozoic Earth Scenario and Log(VMR$_{N_2O})=-3$}
		\tablehead{
               & & \multicolumn{3}{c}{Long Wavelength Cutoff} \\
			\colhead{Parameter} & \colhead{Input} & \colhead{1.8$\mu$m} & \colhead{1.4$\mu$m} & \colhead{1.2$\mu$m}
   }
		\startdata
		\underline{Clouds (Log)} & & & \\
         $P_{\mathrm{top},H_2O}$ (Pa) & 4.85  & $4.5^{+0.22}_{-0.24}$ & $4.43^{+0.27}_{-0.33}$ & $3.35^{+0.97}_{-0.79}$\\
         $D_{H_2O}$ (Pa) & 4.3 & $4.10^{+0.25}_{-0.39}$ & $4.16^{+0.25}_{-0.34}$ & $3.96^{+0.27}_{-0.30}$\\
         $CR_{H_2O}$ & $-3.0$ & $-3.76^{+1.07}_{-2.13}$ & $-3.85^{+1.26}_{-2.08}$ & $-4.51^{+1.70}_{-1.61}$\\
         \underline{Mol. (Log)} & & & &\\
         VMR$_{H_2O}$ & $-2.0$ & $-1.69^{+0.22}_{-0.20}$ & $-1.74^{+0.26}_{-0.21}$ & $-1.00^{+0.44}_{-0.50}$\\
         VMR$_{CH_4}$ & $-4.3$ & $-4.07^{+0.22}_{-0.22}$ & $-3.93^{+0.28}_{-0.25}$ & $-3.50^{+0.37}_{-0.62}$\\
         VMR$_{CO_2}$ & $-3.4$ & $-4.21^{+1.42}_{-1.91}$ & $-3.85^{+1.67}_{-2.08}$ & $-1.54^{+1.47}_{-3.84}$\\
         VMR$_{N_2O}$ & $-3.0$ & $-2.51^{+0.37}_{-0.36}$ & $-1.93^{+0.43}_{-0.44}$ & $-4.10^{+2.05}_{-1.99}$\\
         VMR$_{O_2}$ & $-2.7$ & $-2.58^{+0.66}_{-2.56}$& $-1.88^{+0.57}_{-0.67}$ & $-3.89^{+1.85}_{-2.13}$\\
         VMR$_{O_3}$ & $-6.3$ & $-6.01^{+0.18}_{-0.17}$& $-5.97^{+0.24}_{-0.21}$ & $-5.32^{+0.30}_{-0.54}$\\   
         VMR$_{N_2}$ & $-0.01$ & $-0.01^{+0.01}_{-0.01}$& $-0.02^{+0.01}_{-0.03}$ & $-0.20^{+0.19}_{-4.98}$\\
         \underline{Misc. (Log)} & & & \\
         $A_g$ & $-0.7$ & $-1.18^{+0.60}_{-0.55}$ & $-1.18^{+0.58}_{-0.55}$ & $-1.10^{+0.59}_{-0.57}$\\
         g & 2.99 & $2.99^{+0.01}_{-0.01}$ & $2.99^{+0.01}_{-0.02}$ & $3.02^{+0.01}_{-0.01}$
		\enddata
      \label{proterozoic_table_-3}
       \tablecomments{Input values from \citealt{krissansen2018dis} and \citealt{damiano2023}. }
      \end{deluxetable}   

        \begin{deluxetable}{crrrr}
		\tablecaption{Retrieval Results for Proterozoic Earth Scenario and Log(VMR$_{N_2O})=-4$}
		\tablehead{
               & & \multicolumn{3}{c}{Long Wavelength Cutoff} \\
			\colhead{Parameter} & \colhead{Input} & \colhead{1.8$\mu$m} & \colhead{1.4$\mu$m} & \colhead{1.2$\mu$m}
   }
		\startdata
		\underline{Clouds (Log)} & & & \\
         $P_{\mathrm{top},H_2O}$ (Pa) & 4.85  & $4.33^{+0.23}_{-0.30}$ & $2.78^{+0.41}_{-0.41}$ & $3.79^{+0.47}_{-0.78}$\\
         $D_{H_2O}$ (Pa) & 4.3 & $4.01^{+0.26}_{-0.38}$ & $3.70^{+0.15}_{-0.19}$ & $3.79^{+0.28}_{-0.38}$\\
         $CR_{H_2O}$ & $-3.0$ & $-3.05^{+0.69}_{-2.32}$ & $-4.61^{+1.79}_{-1.54}$ & $-4.32^{+1.28}_{-1.53}$\\
         \underline{Mol. (Log)} & & & &\\
         VMR$_{H_2O}$ & $-2.0$ & $-1.55^{+0.22}_{-0.20}$ & $-0.80^{+0.21}_{-0.17}$ & $-1.03^{+0.53}_{-0.37}$\\
         VMR$_{CH_4}$ & $-4.3$ & $-3.92^{+0.25}_{-0.24}$ & $-2.89^{+0.15}_{-0.16}$ & $-3.90^{+0.53}_{-2.91}$\\
         VMR$_{CO_2}$ & $-3.4$ & $-7.14^{+3.23}_{-3.12}$ & $-0.09^{+0.03}_{-0.06}$ & $-6.13^{+5.11}_{-3.87}$\\
         VMR$_{N_2O}$ & $-4.0$ & $-6.89^{+3.31}_{-3.12}$ & $-6.72^{+3.54}_{-3.21}$ & $-6.52^{+3.77}_{-3.50}$\\
         VMR$_{O_2}$ & $-2.7$ & $-6.61^{+3.42}_{-3.47}$& $-6.43^{+4.11}_{-3.43}$ & $-6.52^{+3.77}_{-3.50}$\\
         VMR$_{O_3}$ & $-6.3$ & $-5.91^{+0.19}_{-0.18}$& $-5.01^{+0.11}_{-0.11}$ & $-5.56^{+0.38}_{-0.35}$\\   
         VMR$_{N_2}$ & $-0.01$ & $-0.01^{+0.01}_{-0.01}$& $-6.35^{+3.63}_{-3.43}$ & $-0.05^{+0.03}_{-4.14}$\\
         \underline{Misc. (Log)} & & & \\
         $A_g$ & $-0.7$ & $-1.20^{+0.58}_{-0.52}$ & $-1.18^{+0.57}_{-0.53}$ & $-1.07^{+0.56}_{-0.57}$\\
         g & 2.99 & $3.00^{+0.01}_{-0.01}$ & $2.99^{+0.01}_{-0.01}$ & $3.02^{+0.01}_{-0.01}$
		\enddata
      \label{proterozoic_table_-4}
       \tablecomments{Input values from \citealt{krissansen2018dis} and \citealt{damiano2023}. }
      \end{deluxetable}  
    
    \section{Discussion and Future Direction} \label{sec:discussion}

    We have shown that even though the detection of CH$_4$ and CO$_2$ in an Archean Earth analog can be made robustly with reflection spectroscopy that extends to near-infrared wavelengths, it will be challenging to address the potential false positive scenarios for the CH$_4$/CO$_2$ biosignature pair with observations in the same wavelength range. Detecting and constraining CO abundance in the atmosphere would help retire the risk of false positives that CH$_4$ came from volcanic outgassing, but the spectral feature of CO in wavelengths $<1.8$ $\mu$m is weak and often buried by stronger absorption of CO$_2$. Here we have considered a generous upper limit of 10\% for CO, and yet the retrieval exercise did not indicate that a useful constraint of the CO abundance would be obtained.
    
    We also performed retrievals for varying wavelength coverage in addition to the full 0.25-1.8$\mu$m range: we cut the long wavelength at 1.7$\mu$m and 1.6$\mu$m and found that CH$_4$ is clearly detected, but the dominant gas in the atmosphere is not correctly identified, resulting in a slight bias in the constraints of the mixing ratio of CH$_4$. The uncertainty of the atmospheric context adds to the ambiguity of the source of the potential biosignature gas.

To test the sensitivity to the assumed CO$_2$ abundance, we additionally simulated a retrieval scenario with a CO$_2$ abundance similar to modern Earth (VMR$_{CO_2}$ = 0.0004, compared to 10\% assumed in the Archean-Earth-like scenario). With 10\% atmospheric CO as the input, the retrieval suggests a CO dominated atmosphere, similar to the results for the 1.6$\mu$m long wavelength cutoff case. Thus, even if we uncover the CO signal by weakening the CO$_2$ feature, we still do not obtain a reliable constraint on CO abundance. Since CO and N$_2$ have the same molecular mass and few spectral features, a high amount of input CO such as 10\% results in degeneracies between the two and the retrieval cannot reliably choose the dominant atmospheric gas species. 

Extending the long wavelength coverage beyond $\sim2$ $\mu$m would play an essential role for interpreting the CH$_4$/CO$_2$ detections. There are stronger and more plentiful CO absorption lines further out in the near-IR and mid-IR such as the ones at $\sim2.4$ $\mu$m and $\sim4.7$ $\mu$m. These features should be explored in future work, along with the impact of thermal emission that becomes more significant at these longer wavelengths, because one might consider preparing the capabilities of starlight suppression and precision photometry/spectroscopy at these longer wavelengths should favorable planets be discovered by the Habitable Worlds Observatory \citep{martin2022next}. 

On the other hand, we found that there is the potential to detect N$_2$O in a Proterozoic Earth atmosphere for relatively large amounts such as Log(VMR$_{N_2O}$) $=-3$ and the appropriate wavelength coverage, whereas smaller abundances like Log(VMR$_{N_2O}$) $=-4$ is potentially non-detectable even with a wide wavelength coverage. Log(VMR$_{N_2O}$) $=-3$ represents an upper limit of N$_2$O abundance which can accumulate in an atmosphere of a planet with a large N$_2$O-producing biomass around a K-type star. This implies that more moderate amounts of N$_2$O will not be detectable, but this should be explored in further studies with an emphasis on the effect of SNR and spectral resolution. Even so, N$_2$O remains a top biosignature candidate in the context of characterizing terrestrial exoplanets with the Habitable Worlds Observatory.

   Another direction to take this study is to consider additional biosignatures potentially present in both in the Archean and Proterozoic eons. Some of the most popular molecules include dimethyl sulfide (DMS) and methyl chloride (CH$_3$Cl), which are overwhelmingly produced biologically and produce features in the near-IR \citep{seager2016}. Reviewing these and other molecular gases will give a more complete picture of the potential to detect biosignatures with future missions.

	\section{Conclusion}
	\label{sec:conclusion}
	
	With efforts ramping up to design the Habitable Worlds Observatory, it is critical to understand the measurements needed to detect biosignatures through direct imaging of Earth-like planets. In this study, we consider exoplanets that may represent Earth in different historical ages. The atmosphere in the Archean eon had negligible amounts of oxygen but was abundant in methane and carbon dioxide. This CH$_4$/CO$_2$ pair can be interpreted as a disequilibrium biosignature, and we showed that each of these molecules would be readily detectable through reflection spectroscopy of an Archean Earth twin exoplanet using the spectral resolution, wavelength range, and SNR currently considered for HWO. We have also shown that it will be challenging to detect the CO molecule, which could rule out the false positive scenario that the CH$_4$ is produced by volcanic outgassing.

    Another alternative biosignature we explored is N$_2$O in a Proterozoic Earth context since the abundance of N$_2$O throughout Earth's history may have been the highest during the Proterozoic eon. N$_2$O alone is a compelling biosignature due to minor known abiotic sources, and we have shown that it may be a promising molecule to pursue due its detectability especially for planets around K-type stars. Further studies could focus on determining the combination of mixing ratio and instrument capabilities needed to constrain the N$_2$O biosignature along with atmospheric context on an exoplanet, such as the relative abundances of other atmospheric gases and instrument SNR/resolution. This work, along with joint studies and future directions, will help inform on the ultimate design and scope of the Habitable Worlds Observatory.
	
	\section*{Acknowledgments}
	We thank Edwin Kite for helpful discussion on false positive scenarios of CH$_4$. This work was supported in part by a strategic initiative of Jet Propulsion Laboratory to improve the understanding of the measurements needed for future astrophysics flagship missions. The High Performance Computing resources used in this investigation were provided by funding from the JPL Information and Technology Solutions Directorate. This research was carried out at the Jet Propulsion Laboratory, California Institute of Technology, under a contract with the National Aeronautics and Space Administration. 
	
	{	\small
		\bibliographystyle{apj}
		\bibliography{bib.bib}

\begin{thebibliography}{}
\expandafter\ifx\csname natexlab\endcsname\relax\def\natexlab#1{#1}\fi
\providecommand{\url}[1]{\href{#1}{#1}}
\providecommand{\dodoi}[1]{doi:~\href{http://doi.org/#1}{\nolinkurl{#1}}}
\providecommand{\doeprint}[1]{\href{http://ascl.net/#1}{\nolinkurl{http://ascl.net/#1}}}
\providecommand{\doarXiv}[1]{\href{https://arxiv.org/abs/#1}{\nolinkurl{https://arxiv.org/abs/#1}}}

\bibitem[{{Airapetian} {et~al.}(2016){Airapetian}, {Glocer}, {Gronoff}, {H{\'e}brard}, \& {Danchi}}]{airapetian2016}
{Airapetian}, V.~S., {Glocer}, A., {Gronoff}, G., {H{\'e}brard}, E., \& {Danchi}, W. 2016, Nature Geoscience, 9, 452, \dodoi{10.1038/ngeo2719}

\bibitem[{{Buessecker} {et~al.}(2022){Buessecker}, {Imanaka}, {Ely}, {Hu}, {Romaniello}, \& {Cadillo-Quiroz}}]{buessecker2022}
{Buessecker}, S., {Imanaka}, H., {Ely}, T., {et~al.} 2022, Nature Geoscience, 15, 1056, \dodoi{10.1038/s41561-022-01089-9}

\bibitem[{{Buick}(2007)}]{buick2007}
{Buick}, R. 2007, Geobiology, 5, 97, \dodoi{10.1111/j.1472-4669.2007.00110.x}

\bibitem[{{Burton} {et~al.}(2013){Burton}, {Sawyer}, \& {Granieri}}]{burton2013}
{Burton}, M.~R., {Sawyer}, G.~M., \& {Granieri}, D. 2013, Reviews in Mineralogy and Geochemistry, 75, 323, \dodoi{10.2138/rmg.2013.75.11}

\bibitem[{Catling \& Zahnle(2020)}]{catling2020archean}
Catling, D.~C., \& Zahnle, K.~J. 2020, Science Advances, 6, eaax1420

\bibitem[{{Catling} {et~al.}(2001){Catling}, {Zahnle}, \& {McKay}}]{catling2001}
{Catling}, D.~C., {Zahnle}, K.~J., \& {McKay}, C.~P. 2001, Science, 293, 839, \dodoi{10.1126/science.1061976}

\bibitem[{{Catling} {et~al.}(2018){Catling}, {Krissansen-Totton}, {Kiang}, {Crisp}, {Robinson}, {DasSarma}, {Rushby}, {Del Genio}, {Bains}, \& {Domagal-Goldman}}]{catling2018}
{Catling}, D.~C., {Krissansen-Totton}, J., {Kiang}, N.~Y., {et~al.} 2018, Astrobiology, 18, 709, \dodoi{10.1089/ast.2017.1737}

\bibitem[{{Chen} {et~al.}(2015){Chen}, {Mothapo}, \& {Shi}}]{chen2015}
{Chen}, H., {Mothapo}, N.~V., \& {Shi}, W. 2015, Microbial Ecology, 69, 180, \dodoi{10.1007/s00248-014-0488-0}

\bibitem[{{Damiano} \& {Hu}(2020)}]{Damiano2020a}
{Damiano}, M., \& {Hu}, R. 2020, \aj, 159, 175, \dodoi{10.3847/1538-3881/ab79a5}

\bibitem[{Damiano \& Hu(2022)}]{damiano2022}
Damiano, M., \& Hu, R. 2022, \aj, 163, 299, \dodoi{10.3847/1538-3881/ac6b97}

\bibitem[{{Damiano} {et~al.}(2023){Damiano}, {Hu}, \& {Mennesson}}]{damiano2023}
{Damiano}, M., {Hu}, R., \& {Mennesson}, B. 2023, \aj, 166, 157, \dodoi{10.3847/1538-3881/acefd3}

\bibitem[{{Des Marais} {et~al.}(2002){Des Marais}, {Harwit}, {Jucks}, {Kasting}, {Lin}, {Lunine}, {Schneider}, {Seager}, {Traub}, \& {Woolf}}]{desmarais2002}
{Des Marais}, D.~J., {Harwit}, M.~O., {Jucks}, K.~W., {et~al.} 2002, Astrobiology, 2, 153, \dodoi{10.1089/15311070260192246}

\bibitem[{Fakhraee {et~al.}(2023)Fakhraee, Tarhan, Reinhard, Crowe, Lyons, \& Planavsky}]{fakhraee2023}
Fakhraee, M., Tarhan, L.~G., Reinhard, C.~T., {et~al.} 2023, Earth-Science Reviews, 240, 104398, \dodoi{https://doi.org/10.1016/j.earscirev.2023.104398}

\bibitem[{{Fujii} {et~al.}(2018){Fujii}, {Angerhausen}, {Deitrick}, {Domagal-Goldman}, {Grenfell}, {Hori}, {Kane}, {Pall{\'e}}, {Rauer}, {Siegler}, {Stapelfeldt}, \& {Stevenson}}]{fujii2018}
{Fujii}, Y., {Angerhausen}, D., {Deitrick}, R., {et~al.} 2018, Astrobiology, 18, 739, \dodoi{10.1089/ast.2017.1733}

\bibitem[{{Gao} {et~al.}(2015){Gao}, {Hu}, {Robinson}, {Li}, \& {Yung}}]{gao2015}
{Gao}, P., {Hu}, R., {Robinson}, T.~D., {Li}, C., \& {Yung}, Y.~L. 2015, \apj, 806, 249, \dodoi{10.1088/0004-637X/806/2/249}

\bibitem[{{Gaudi} {et~al.}(2020){Gaudi}, {Seager}, {Mennesson}, {Kiessling}, {Warfield}, {Cahoy}, {Clarke}, {Domagal-Goldman}, {Feinberg}, {Guyon}, {Kasdin}, {Mawet}, {Plavchan}, {Robinson}, {Rogers}, {Scowen}, {Somerville}, {Stapelfeldt}, {Stark}, {Stern}, {Turnbull}, {Amini}, {Kuan}, {Martin}, {Morgan}, {Redding}, {Stahl}, {Webb}, {Alvarez-Salazar}, {Arnold}, {Arya}, {Balasubramanian}, {Baysinger}, {Bell}, {Below}, {Benson}, {Blais}, {Booth}, {Bourgeois}, {Bradford}, {Brewer}, {Brooks}, {Cady}, {Caldwell}, {Calvet}, {Carr}, {Chan}, {Cormarkovic}, {Coste}, {Cox}, {Danner}, {Davis}, {Dewell}, {Dorsett}, {Dunn}, {East}, {Effinger}, {Eng}, {Freebury}, {Garcia}, {Gaskin}, {Greene}, {Hennessy}, {Hilgemann}, {Hood}, {Holota}, {Howe}, {Huang}, {Hull}, {Hunt}, {Hurd}, {Johnson}, {Kissil}, {Knight}, {Kolenz}, {Kraus}, {Krist}, {Li}, {Lisman}, {Mandic}, {Mann}, {Marchen}, {Marrese-Reading}, {McCready}, {McGown}, {Missun}, {Miyaguchi}, {Moore}, {Nemati}, {Nikzad}, {Nissen}, {Novicki}, {Perrine}, {Pineda}, {Polanco},
  {Putnam}, {Qureshi}, {Richards}, {Eldorado Riggs}, {Rodgers}, {Rud}, {Saini}, {Scalisi}, {Scharf}, {Schulz}, {Serabyn}, {Sigrist}, {Sikkia}, {Singleton}, {Shaklan}, {Smith}, {Southerd}, {Stahl}, {Steeves}, {Sturges}, {Sullivan}, {Tang}, {Taras}, {Tesch}, {Therrell}, {Tseng}, {Valente}, {Van Buren}, {Villalvazo}, {Warwick}, {Webb}, {Westerhoff}, {Wofford}, {Wu}, {Woo}, {Wood}, {Ziemer}, {Arney}, {Anderson}, {Ma{\'\i}z-Apell{\'a}niz}, {Bartlett}, {Belikov}, {Bendek}, {Cenko}, {Douglas}, {Dulz}, {Evans}, {Faramaz}, {Feng}, {Ferguson}, {Follette}, {Ford}, {Garc{\'\i}a}, {Geha}, {Gelino}, {G{\"o}tberg}, {Hildebrand t}, {Hu}, {Jahnke}, {Kennedy}, {Kreidberg}, {Isella}, {Lopez}, {Marchis}, {Macri}, {Marley}, {Matzko}, {Mazoyer}, {McCandliss}, {Meshkat}, {Mordasini}, {Morris}, {Nielsen}, {Newman}, {Petigura}, {Postman}, {Reines}, {Roberge}, {Roederer}, {Ruane}, {Schwieterman}, {Sirbu}, {Spalding}, {Teplitz}, {Tumlinson}, {Turner}, {Werk}, {Wofford}, {Wyatt}, {Young}, \& {Zellem}}]{Gaudi2020}
{Gaudi}, B.~S., {Seager}, S., {Mennesson}, B., {et~al.} 2020, arXiv e-prints, arXiv:2001.06683.
\newblock \doarXiv{2001.06683}

\bibitem[{{Hall} {et~al.}(2023){Hall}, {Krissansen-Totton}, {Robinson}, {Salvador}, \& {Fortney}}]{hall2023}
{Hall}, S., {Krissansen-Totton}, J., {Robinson}, T., {Salvador}, A., \& {Fortney}, J.~J. 2023, \aj, 166, 254, \dodoi{10.3847/1538-3881/ad03e9}

\bibitem[{{Hargreaves} {et~al.}(2019){Hargreaves}, {Gordon}, {Rothman}, {Tashkun}, {Perevalov}, {Lukashevskaya}, {Yurchenko}, {Tennyson}, \& {M{\"u}ller}}]{hargreaves2019}
{Hargreaves}, R.~J., {Gordon}, I.~E., {Rothman}, L.~S., {et~al.} 2019, \jqsrt, 232, 35, \dodoi{10.1016/j.jqsrt.2019.04.040}

\bibitem[{{Harman} {et~al.}(2015){Harman}, {Schwieterman}, {Schottelkotte}, \& {Kasting}}]{harman2015}
{Harman}, C.~E., {Schwieterman}, E.~W., {Schottelkotte}, J.~C., \& {Kasting}, J.~F. 2015, \apj, 812, 137, \dodoi{10.1088/0004-637X/812/2/137}

\bibitem[{{Hu} \& {Delgado Diaz}(2019)}]{Hu2019}
{Hu}, R., \& {Delgado Diaz}, H. 2019, \apj, 886, 126, \dodoi{10.3847/1538-4357/ab4cea}

\bibitem[{{Kasting}(2005)}]{kasting2005}
{Kasting}, J. 2005, Precambrian Research, 137, 119, \dodoi{10.1016/j.precamres.2005.03.002}

\bibitem[{Kharecha {et~al.}(2005)Kharecha, Kasting, \& Sierfert}]{kharecha2005}
Kharecha, P., Kasting, J., \& Sierfert, J. 2005, Geobiology, 3, 53, \dodoi{https://doi.org/10.1111/j.1472-4669.2005.00049.x}

\bibitem[{Knowles(1982)}]{knowles1982}
Knowles, R. 1982, Microbiological reviews, 46, 43, \dodoi{10.1128/mr.46.1.43-70.1982}

\bibitem[{{Krissansen-Totton} {et~al.}(2018){Krissansen-Totton}, {Olson}, \& {Catling}}]{krissansen2018dis}
{Krissansen-Totton}, J., {Olson}, S., \& {Catling}, D.~C. 2018, Science Advances, 4, eaao5747, \dodoi{10.1126/sciadv.aao5747}

\bibitem[{{Leger} {et~al.}(1993){Leger}, {Pirre}, \& {Marceau}}]{leger1993}
{Leger}, A., {Pirre}, M., \& {Marceau}, F.~J. 1993, \aap, 277, 309

\bibitem[{Lemke {et~al.}(2007)Lemke, Ren, Alley, Allison, Carrasco, Flato, Fujii, Kaser, Mote, Thomas, \& Zhang}]{lemke2007}
Lemke, P., Ren, J., Alley, R., {et~al.} 2007, IPCC, 2007. Climate Change 2007. Synthesis Report. Contribution of Working Groups I, II \& III to the Fourth Assessment Report of the Intergovernmental Panel on Climate Change. Geneva (Cambridge University Press,)

\bibitem[{{Luger} \& {Barnes}(2015)}]{luger2015}
{Luger}, R., \& {Barnes}, R. 2015, Astrobiology, 15, 119, \dodoi{10.1089/ast.2014.1231}

\bibitem[{{Lyons} {et~al.}(2014){Lyons}, {Reinhard}, \& {Planavsky}}]{lyons2014}
{Lyons}, T.~W., {Reinhard}, C.~T., \& {Planavsky}, N.~J. 2014, \nat, 506, 307, \dodoi{10.1038/nature13068}

\bibitem[{{Mamajek} \& {Stapelfeldt}(2024)}]{mamajek2024}
{Mamajek}, E., \& {Stapelfeldt}, K. 2024, arXiv e-prints, arXiv:2402.12414, \dodoi{10.48550/arXiv.2402.12414}

\bibitem[{Martin {et~al.}(2022)Martin, Lawrence, Redding, Mennesson, Rodgers, Hurd, Morgan, Hu, Steeves, Jewell, {et~al.}}]{martin2022next}
Martin, S., Lawrence, C., Redding, D., {et~al.} 2022, Journal of Astronomical Telescopes, Instruments, and Systems, 8, 044005

\bibitem[{{Meadows} {et~al.}(2018){Meadows}, {Reinhard}, {Arney}, {Parenteau}, {Schwieterman}, {Domagal-Goldman}, {Lincowski}, {Stapelfeldt}, {Rauer}, {DasSarma}, {Hegde}, {Narita}, {Deitrick}, {Lustig-Yaeger}, {Lyons}, {Siegler}, \& {Grenfell}}]{meadows2018}
{Meadows}, V.~S., {Reinhard}, C.~T., {Arney}, G.~N., {et~al.} 2018, Astrobiology, 18, 630, \dodoi{10.1089/ast.2017.1727}

\bibitem[{Meadows {et~al.}(2018)Meadows, Reinhard, Arney, Parenteau, Schwieterman, Domagal-Goldman, Lincowski, Stapelfeldt, Rauer, DasSarma, {et~al.}}]{meadows2018exoplanet}
Meadows, V.~S., Reinhard, C.~T., Arney, G.~N., {et~al.} 2018, Astrobiology, 18, 630

\bibitem[{Meister \& Reyes(2019)}]{meister2019}
Meister, P., \& Reyes, C. 2019, Geosciences, 9, \dodoi{10.3390/geosciences9120507}

\bibitem[{{Miyazaki} \& {Korenaga}(2022)}]{miyazaki2022}
{Miyazaki}, Y., \& {Korenaga}, J. 2022, \nat, 603, 86, \dodoi{10.1038/s41586-021-04371-9}

\bibitem[{National Academies~of Sciences {et~al.}(2021)}]{national2021pathways}
National Academies~of Sciences, Engineering, M., {et~al.} 2021

\bibitem[{{Ohtomo} {et~al.}(2014){Ohtomo}, {Kakegawa}, {Ishida}, {Nagase}, \& {Rosing}}]{ohtomo2014}
{Ohtomo}, Y., {Kakegawa}, T., {Ishida}, A., {Nagase}, T., \& {Rosing}, M.~T. 2014, Nature Geoscience, 7, 25, \dodoi{10.1038/ngeo2025}

\bibitem[{Pinto {et~al.}(2021)Pinto, Weigelhofer, Brito, \& Hein}]{pinto2021}
Pinto, R., Weigelhofer, G., Brito, A.~G., \& Hein, T. 2021, PeerJ, 9, e10767, \dodoi{10.7717/peerj.10767}

\bibitem[{Quick {et~al.}(2019)Quick, Reeder, Farrell, Tonina, Feris, \& Benner}]{quick2019}
Quick, A.~M., Reeder, W.~J., Farrell, T.~B., {et~al.} 2019, Earth-Science Reviews, 191, 224, \dodoi{https://doi.org/10.1016/j.earscirev.2019.02.021}

\bibitem[{Ragsdale(2004)}]{ragsdale2004}
Ragsdale, S.~W. 2004, Critical reviews in biochemistry and molecular biology, 39, 165, \dodoi{10.1080/10409230490496577}

\bibitem[{{Rauer} {et~al.}(2011){Rauer}, {Gebauer}, {Paris}, {Cabrera}, {Godolt}, {Grenfell}, {Belu}, {Selsis}, {Hedelt}, \& {Schreier}}]{rauer2011}
{Rauer}, H., {Gebauer}, S., {Paris}, P.~V., {et~al.} 2011, \aap, 529, A8, \dodoi{10.1051/0004-6361/201014368}

\bibitem[{{Rothman} {et~al.}(2010){Rothman}, {Gordon}, {Barber}, {Dothe}, {Gamache}, {Goldman}, {Perevalov}, {Tashkun}, \& {Tennyson}}]{rothman2010}
{Rothman}, L.~S., {Gordon}, I.~E., {Barber}, R.~J., {et~al.} 2010, \jqsrt, 111, 2139, \dodoi{10.1016/j.jqsrt.2010.05.001}

\bibitem[{{Sagan} {et~al.}(1993){Sagan}, {Thompson}, {Carlson}, {Gurnett}, \& {Hord}}]{sagan1993}
{Sagan}, C., {Thompson}, W.~R., {Carlson}, R., {Gurnett}, D., \& {Hord}, C. 1993, \nat, 365, 715, \dodoi{10.1038/365715a0}

\bibitem[{{Samarkin} {et~al.}(2010){Samarkin}, {Madigan}, {Bowles}, {Casciotti}, {Priscu}, {McKay}, \& {Joye}}]{samarkin2010}
{Samarkin}, V.~A., {Madigan}, M.~T., {Bowles}, M.~W., {et~al.} 2010, Nature Geoscience, 3, 341, \dodoi{10.1038/ngeo847}

\bibitem[{{Schindler} \& {Kasting}(2000)}]{schindler2000}
{Schindler}, T.~L., \& {Kasting}, J.~F. 2000, \icarus, 145, 262, \dodoi{10.1006/icar.2000.6340}

\bibitem[{{Schumann} \& {Huntrieser}(2007)}]{schumann2007}
{Schumann}, U., \& {Huntrieser}, H. 2007, Atmospheric Chemistry \& Physics, 7, 3823, \dodoi{10.5194/acp-7-3823-200710.5194/acpd-7-2623-2007}

\bibitem[{{Schwieterman} {et~al.}(2019){Schwieterman}, {Reinhard}, {Olson}, {Ozaki}, {Harman}, {Hong}, \& {Lyons}}]{schweiterman2019}
{Schwieterman}, E.~W., {Reinhard}, C.~T., {Olson}, S.~L., {et~al.} 2019, \apj, 874, 9, \dodoi{10.3847/1538-4357/ab05e1}

\bibitem[{Schwieterman {et~al.}(2018)Schwieterman, Kiang, Parenteau, Harman, DasSarma, Fisher, Arney, Hartnett, Reinhard, Olson, {et~al.}}]{schwieterman2018exoplanet}
Schwieterman, E.~W., Kiang, N.~Y., Parenteau, M.~N., {et~al.} 2018, Astrobiology, 18, 663

\bibitem[{{Schwieterman} {et~al.}(2022){Schwieterman}, {Olson}, {Pidhorodetska}, {Reinhard}, {Ganti}, {Fauchez}, {Bastelberger}, {Crouse}, {Ridgwell}, \& {Lyons}}]{schwieterman2022}
{Schwieterman}, E.~W., {Olson}, S.~L., {Pidhorodetska}, D., {et~al.} 2022, \apj, 937, 109, \dodoi{10.3847/1538-4357/ac8cfb}

\bibitem[{{Seager} {et~al.}(2016){Seager}, {Bains}, \& {Petkowski}}]{seager2016}
{Seager}, S., {Bains}, W., \& {Petkowski}, J.~J. 2016, Astrobiology, 16, 465, \dodoi{10.1089/ast.2015.1404}

\bibitem[{{Segura} {et~al.}(2005){Segura}, {Kasting}, {Meadows}, {Cohen}, {Scalo}, {Crisp}, {Butler}, \& {Tinetti}}]{segura2005}
{Segura}, A., {Kasting}, J.~F., {Meadows}, V., {et~al.} 2005, Astrobiology, 5, 706, \dodoi{10.1089/ast.2005.5.706}

\bibitem[{{Sholes} {et~al.}(2019){Sholes}, {Krissansen-Totton}, \& {Catling}}]{sholes2019}
{Sholes}, S.~F., {Krissansen-Totton}, J., \& {Catling}, D.~C. 2019, Astrobiology, 19, 655, \dodoi{10.1089/ast.2018.1835}

\bibitem[{{The LUVOIR Team}(2019)}]{luvoir2019}
{The LUVOIR Team}. 2019, arXiv e-prints, arXiv:1912.06219, \dodoi{10.48550/arXiv.1912.06219}

\bibitem[{{Thompson} {et~al.}(2022){Thompson}, {Krissansen-Totton}, {Wogan}, {Telus}, \& {Fortney}}]{thompson2022}
{Thompson}, M.~A., {Krissansen-Totton}, J., {Wogan}, N., {Telus}, M., \& {Fortney}, J.~J. 2022, Proceedings of the National Academy of Science, 119, e2117933119, \dodoi{10.1073/pnas.2117933119}

\bibitem[{{Wogan} {et~al.}(2020){Wogan}, {Krissansen-Totton}, \& {Catling}}]{wogan2020}
{Wogan}, N., {Krissansen-Totton}, J., \& {Catling}, D.~C. 2020, The Planetary Science Journal, 1, 58, \dodoi{10.3847/PSJ/abb99e}

\end{thebibliography}
	}

\end{document}